\documentclass{amsart}

\title{A toy model for the W/Z mass ratio}
\author{Robert Arnott Wilson}
\date{17th April 2022}

\address{Queen Mary University of London}

\email{r.a.wilson@qmul.ac.uk}
\newcommand{\RR}{\mathbf R}
\newcommand{\CC}{\mathbf C}

\begin{document}
\begin{abstract}
The recently reported $7\sigma$ anomaly in measurements of the W/Z mass ratio, if confirmed,
will demand an extension of some kind to the standard model of particle physics.
In this paper I consider whether some recently proposed models are capable, in principle,
of resolving this anomaly, subject to experimental confirmation.
\end{abstract}
\maketitle

\section{Introduction}
The announcement \cite{WZ} on 8th April 2022 of an anomalous measurement of the W/Z mass ratio, some $.1\%$ higher than the
standard model prediction, is the strongest signal yet that the standard model of particle physics (SMPP)
is missing some important ingredient. Of course, it is
possible that there is some unexplained source of error in either the theoretical prediction, or the experiment and its analysis, or both,
that resolves this issue without 
change to the standard model itself. However, it is also timely to consider whether some `new'
or beyond the standard model (BSM) physics may be responsible. Here I consider two recently proposed models
\cite{octions,ZYM} to see whether they contain ingredients that could, at least in principle, explain the anomaly.

The first of these \cite{octions} 
is a Grand Unified Theory (GUT) based on the Lie group $Spin(12,4)$ of dimension $120$ as the overarching gauge group, split into
the Lorentz group $Spin(3,1)$ and other pieces from which the standard model gauge groups are derived. The group $Spin(12,4)$
is embedded in a Lie group of type $E_8$, purely for the purpose of uniting spinor and vector representations into a single structure.
Other than providing a convenient notation \cite{E8}, the full structure of $E_8$ is not really used (so far) in this model.

The second model \cite{ZYM}
is based not on a Lie group of dimension $120$, but on a finite group of order $120$. The motivation here is to use the
discrete structure for fundamental physics \cite{tHooft22,tHooft17} 
at the Planck scale (or whatever scale turns out to be most appropriate), in order to avoid the
ultraviolet (UV) catastrophe that afflicts all known continuous theories \cite{Baez}. At practical scales, of course, the discrete structure is
embedded in continuous structures that provide all the necessary quantum field theories (QFTs) of the SMPP, and classical fields,
for practical calculations \cite{icosa}.

Although there are similarities between the two models in certain aspects, they are quite different in others. The first is a special relativistic
model, as it contains a Lorentz group $Spin(3,1)$ but not $GL(4,\RR)$. The second does contain $GL(4,\RR)$, and is therefore
potentially a generally covariant model (although this has not been explicitly proved). If so, then at the discrete level it must contain
a quantum theory of gravity of some kind. If we are interested in BSM physics, therefore, this second model is much more likely to
provide the necessary ingredients.

\section{Einstein mass and Dirac mass}
A peculiarity of the second model that is not shared by the first model, or by any other conventional GUT, is that the map between the
Lorentz group in the Dirac form $SL(2,\CC)$ and in the Einstein form $SO(3,1)$ is only determined locally. 
It is this property that allows the possibility of general covariance. Since it is the Lorentz group that
defines the concept of rest mass, this implies that, in this model, rest masses of elementary particles can in principle only be defined locally, by a local
calibration of the Dirac rest mass against the Einstein rest mass. If an experiment takes place `non-locally', either in space or in time,
or both, then it
is necessary to consider to what extent the concept of `rest mass' remains the same, as the concept of `rest' varies.

Roughly speaking, although not precisely, Dirac mass corresponds to inertial mass and passive gravitational mass, which are known 
experimentally to be indistinguishable to very high accuracy, for macroscopic objects. Einstein mass roughly corresponds to active
gravitational mass, which cannot be definitively distinguished from inertial mass either, but here the experimental relative uncertainty is hard to
control below about $10^{-4}$, or $10^{-5}$ at best. Moreover, these figures apply to macroscopic objects only, and there is no theoretical or
experimental guarantee that they can be extrapolated over the 20 or 30 orders of magnitude (or more) required to apply them to elementary particles.
Therefore we cannot assume without proof that Dirac masses and Einstein masses of elementary particles transform in the same way
under non-inertial coordinate transformations.

In principle, masses of (elementary) fermions are, by definition, Dirac masses, while masses of 
(elementary) bosons are Einstein masses. In particle physics, masses
of fermions such as the electron, proton and muon are known very precisely, while masses of bosons such as the W and Z are known
much less precisely. In practice, therefore, mass is always defined as Dirac mass. Measuring the Dirac mass of a boson always involves
the escape of a neutrino or anti-neutrino, and it is the measurement of the energies of these neutrinos that constitutes one of the main
difficulties in experimental measurement of the mass of the W boson.

The calibration of Einstein and Dirac masses against each other 
is normally done by comparing masses of electrons and protons (fermions)
against water molecules (bosons), in the context of (originally) classical electrodynamics or (now) quantum electrodynamics (QED).
The calibration produces a single dimensionless constant that defines the Dirac mass of bosons locally, 
or equivalently the Einstein mass of fermions, and this constant can be taken to be the
electron/proton mass ratio. This constant was first fixed to a relative accuracy of $10^{-4}$ in  1969, though 
with significant tension between different measurements \cite{1969} over the period from 1949 to 1967.
This tension later 
became largely irrelevant in the refinement to $10^{-6}$ in the 1973 CODATA recommendations for
fundamental physical constants 
\cite{1973}.

Effectively, therefore, the equivalence of Einstein and Dirac masses that is built into the standard model was defined locally in 1973. The question is,
then, to what extent it is still valid in 2022. Direct experimental measurements of Einstein mass using gravity cannot in practice achieve
a consistent (over all types of materials) accuracy much better than $10^{-4}$. Experiments that claim a better accuracy than this for measurements
of Newton's gravitational constant $G$ are in some cases inconsistent with each other \cite{Gillies,newG,QLi}. 
The accuracy of the calibration of Einstein versus Dirac masses cannot
therefore be assumed to be much better than this.

The experiment under discussion does not claim an accuracy greater than this, however, and reports a relative anomaly of $10^{-3}$. 
Hence this calibration issue cannot be to blame, unless there is something else that affects the W and/or Z bosons, but does not affect
ordinary bosonic matter in the form of atoms or molecules. The obvious extra ingredient is the neutrinos, which are not involved in the
electromagnetic experiments that define the calibrations, but are involved in the weak force experiments that measure the W and Z masses.

\section{Neutrinos}
The biggest of the many problems in the study of neutrinos is the neutrino mass problem. Neutrino oscillations
\cite{oscillation,neutrinos,SNO} imply that at least two flavours of
neutrinos have non-zero mass, but experiment cannot directly measure this mass. The model under discussion potentially resolves this issue by
separating the Dirac mass, which must be non-zero according to the standard model, from the Einstein mass, which as far as experiment
tells us might be exactly zero. In other words, the model implies that neutrinos exhibit a difference between Einstein and Dirac mass, that is
separate from the differences already considered. While we cannot directly measure this difference for individual neutrinos, it is possible
that it can be measured indirectly. In principle, any experiment that involves neutrinos, and therefore measures weak force processes,
might do this.

In other words, we need to consider the possibility that the standard model predicts the Dirac mass of the W boson, and the experiment
under discussion has
measured the Einstein mass, or vice versa. No experiment has ever calibrated Einstein mass against Dirac mass using the W and Z bosons,
so that this is the first experiment that provides a strong signal that these two types of mass may vary with respect to each other
over a period of years. Earlier experimental measurements of the W/Z mass ratio
have indeed found a weak signal of this phenomenon, but not at a 
statistically significant level \cite{WZ}.

In order to test this hypothesis, we need at least a toy model of quantum gravity (QG), to allow the gravitational field to interact with the neutrinos in order
to produce the measured effect. Experimentally, we have absolutely no chance of making direct measurements of gravitational effects on
individual neutrinos. In particular, there is no evidence at all that all neutrinos are affected in the same way by the gravitational field, and
some circumstantial evidence that the three flavours are affected differently. More seriously, we need at least a billion neutrinos to make up
the mass of a single neutron, so that the best we could ever hope for is to measure an average over billions of neutrinos. The effect of
QG on an individual neutrino could therefore be a billion times stronger than the average, without contradicting experiment in any way.

Moreover, neutrinos are known to change flavour as they travel through matter. Travelling through matter
implies travelling through a changing gravitational field. It is therefore reasonable to suppose that the mechanism by which
this flavour change occurs could be a direct interaction with the quantum gravitational field. 
It is not necessary, however, to specify exactly how this interaction occurs. In the standard model, flavour eigenstates
are not the same as mass eigenstates, which in the model under discussion translates into saying that the Einstein (flavour) eigenstates
are not the same as the Dirac (mass) eigenstates.

A full 
QG model of these processes is not available at the present time, but we can still attempt to build a toy model
of the quantum effects of the gravity of the Earth, Moon and Sun on the neutrinos measured in this experiment. Local effects on a
terrestrial scale must average out, so there is little point in considering any possible effects due to the time of day, 
the phase of the Moon, 
the tides,
or the latitude of the experiment. We must assume that the effect is more subtle than this, and manifests itself on
a timescale of years due to longer-term variations such as changes in the angle of tilt of the Earth's axis and/or the angle of inclination of
the Moon's orbit. Shorter timescale effects are hidden by noise, and/or eliminated by averaging and/or by re-calibration of the experiment.

\section{A toy model}
In the standard model, the W/Z mass ratio is usually expressed as $\cos\theta_W$, where $\theta_W$ is the electro-weak mixing angle
(Weinberg angle).
In the discrete model, this angle is now interpreted as a gravity mixing angle, but we are mixing three gravitational fields, so we need at least
two angles. Both these angles have to be mapped into a single copy of the complex numbers in order to match the standard model, so we should
see either the sum or the difference of the two angles. 

Now the reported anomaly translates to the difference between an expected angle
of $28.18^\circ$ and an experimental value of $28.11^\circ$. The toy model is made up of an angle of $23.44^\circ$ 
(the Earth's axial tilt) that varies quite slowly,
at a current average rate of about $-.013^\circ$ per century,
and an angle that varies on a 347 day cycle between $4.99^\circ$ and $5.30^\circ$ (the Moon's orbital inclination).
If we simply add these together we obtain an angle varying between $28.43^\circ$ and $28.74^\circ$.
Clearly this is not exactly correct, and the toy model does not take all the necessary parameters into account. Moreover, there is a significant loss of
information in the reduction from a 3-dimensional geometry to a 2-dimensional geometry, by ignoring the
phase of the Moon. This is however necessary in order to remove a 1-dimensional
gravitational field strength parameter and
reduce to a single dimensionless SMPP parameter. 

The important aspect of the toy model, however, is the prediction that the angle is not constant over time, on the scale of years that this
experiment has been running. If we normalise to the standard model prediction of $28.18^\circ$, then the expected range normalises to
the range $28.03^\circ$ to $28.34^\circ$. Previous experimental measurements of the W/Z mass ratio have indeed spanned approximately this range,
although the experimental uncertainty has been too large to claim any real statistical significance to this variation.
The new reported value of $28.11^\circ$, with a much smaller uncertainty, is however consistent with this toy model, but not with the
standard model.

\section{Further work}
In order to determine whether the extension to the SMPP proposed in \cite{icosa}
really does resolve the anomaly, it will be necessary to do much more detailed
modelling of a $3$-dimensional rather than $2$-dimensional gravitational field, and to analyse the experimental data in the
context of a varying prediction over time, most likely in the range $\pm.14\%$, rather than a single constant value.
It is impossible to say at this stage whether this work is likely to confirm this prediction or falsify it. But it can definitely distinguish between the
standard model and the proposed extension, and should therefore enable us to rule out one or the other, or possibly both.

At the same time, the model makes a similar prediction for the mass ratio of charged to neutral kaons, that depends only on a single angle,
and is therefore much easier to analyse, if anyone cares to do the experiment. The prediction is that this boson mass ratio also varies
with a period of 347 days, in the range $\pm.05\%$, if measured via weak interactions. This variation does not apply to measurements
made by purely electrodynamic interactions, however. It is based on the analysis in \cite{icosa}, in which the kaon mass ratio is
conjectured to be approximately $\cos^2\theta_M$, where $\theta_M$ denotes the inclination of the Moon's orbit.
This prediction is therefore a stringent test of my extended model against the pure standard model.

On the theoretical side, the main challenge is to develop a full quantum theory of gravity on the foundations laid in \cite{ZYM}.
The experimental results suggest that some of the basic assumptions of conventional 
QG theories may not be
valid in the real universe. In particular, if it is confirmed that Einstein mass and Dirac mass transform in different ways under
non-inertial coordinate transformations, as the new model implies, 
then we cannot necessarily assume that the graviton transforms as a spin $2$ Dirac
particle, even if it has spin $2$ in the Einstein sense. 
Experiment suggests that we should look instead at properties of neutrinos, the W and Z bosons, and possibly also kaons.

Indeed, Hossenfelder \cite{hossenfelder} has previously suggested virtual kaons as playing a role in quantum gravity. 
Since a kaon can decay into a pion with all the excess energy going into a neutrino and an antineutrino 
\cite{kaonanomaly}, a neutrino passing through a point in space
has similar combinatorial properties to a virtual kaon (minus a virtual pion) located at that point.
Moreover, this particular decay mode occurs more often than the standard model predicts, suggesting that 
it may indeed hide a 
QG effect that is not recognised by the standard model.

If neutrinos have Dirac mass
but no Einstein mass, then they can travel at the speed of light in a vacuum, carrying information with them, and can therefore
participate in the creation of an `information horizon' as suggested by McCulloch \cite{mcculloch} in his theory of quantum inertia (QI).
It is therefore possible that the model proposed in \cite{ZYM} might provide a deeper theoretical underpinning for these already existing
models of gravity and inertia. If so, then no spin $2$ graviton would be required, and the real work would be done instead by the neutrinos, 
the fundamental fermions that provide the necessary translation between
Einstein (bosonic) mass and Dirac (fermionic) mass.

\section{Conclusion}
The extension to the SMPP proposed in \cite{ZYM} suggests that the 
first place to look for explanations of experimental anomalies is a contamination of the
experiment by 
QG effects. Such effects probably cannot be detected in
ordinary matter, made of electrons, protons and neutrons. This is because the model offers each
observer three scalars, and if these are chosen to be the masses of these three particles, then
the standard model can be built on the assumption that they are the same for all
observers. 
Exotic particles mostly seem to belong to
vector representations, not scalars, so cannot necessarily be assumed to have the same mass for all observers.
Hence it should be expected that quantum gravitational effects would be detectable in
non-local experiments involving 
such particles. 
Possible QG signals 
in the muon $g-2$ anomaly,
and neutrino and kaon oscillations, 
have been discussed in \cite{icosa}. The present paper adds the W mass anomaly to the list.

\end{document}